\documentclass[a4paper,12pt]{article}
\usepackage[T1]{fontenc}   
\usepackage{setspace}

\usepackage{geometry}


\usepackage{lscape,multicol}
\usepackage{epsfig}
\usepackage{graphics}
\usepackage{latexsym}
\usepackage{amssymb}
\usepackage{amsmath}
\usepackage{amsfonts}
\usepackage{amsxtra}

\newcommand{\emu}{e_{\mu}}
\newcommand{\id}{\mathbb{I}}
\newcommand{\ice}[1]{\relax}

\def\tr{\text{Tr}}
\def\re{\mathfrak{Re}}

\def\MFPlat{\mathcal{M}^{\text{lat}}_{FP}}
\date{}
\title{Scaling Properties of the Probability Distribution of Lattice Gribov Copies}
\author{A.Y.~Lokhov$^a$, O. P\`ene$^b$, C.~Roiesnel$^a$}
\begin{document}
\maketitle
\begin{center}
$^a$ Centre de Physique Th\'eorique\footnote{
Unit\'e Mixte de Recherche 7644 du Centre National de
la Recherche Scientifique} de l'\'Ecole polytechnique\\
F91128 Palaiseau cedex, France\\
$^b$Laboratoire de Physique Th\'eorique et Hautes
Energies\footnote{Unit\'e Mixte de Recherche 8627 du Centre National de
la Recherche Scientifique}\\
{Universit\'e de Paris XI, B\^atiment 211, 91405 Orsay Cedex,
France}\\
\end{center}
\begin{abstract}
We study the problem of the Landau gauge fixing in the case of the  $SU(2)$
lattice gauge theory. We show that the probability to find a lattice Gribov
copy increases considerably when the physical size of the lattice exceeds
some critical value $\approx2.75/\sqrt{\sigma}$, almost independent of the
lattice spacing. The impact of the choice of the copy on Green functions is
presented. We confirm that the ghost propagator depends on the choice of the
copy, this dependence  decreasing for increasing volumes above the
critical one. The gluon propagator as well as the gluonic three-point
 functions are insensitive to choice of the copy (within present
statistical errors). Finally we show that gauge copies which have the same value of the
minimisation functional ($\int d^4x (A^a_\mu)^2$) are equivalent, up to a
global gauge transformation,  and yield the same Green functions. 
\end{abstract}
{\begin{flushright}
{\small CPHT-RR 065.1105}\\
{\small LPT-Orsay/05-76}\\
\end{flushright}
%
\section{Introduction}
%
%

The question of the quantisation in the infrared of a non-Abelian gauge
theory is very important for the understanding of diverse non-perturbative
phenomena. The main problem  that arises while performing the quantisation
in the covariant gauge is the problem of the  non-uniqueness of the solution
of the equation specifying the gauge-fixing condition, that was pointed out
by  Gribov~\cite{Gribov:1977wm}. In fact, the solution is unique for fields
having small magnitude. A solution proposed in \cite{Gribov:1977wm} consists
in restricting the (functional ) integration over gauge fields in the
partition function to the region inside the so  called Gribov horizon (the
surface, closest to the trivial field $\mathcal{A}_\mu=0$,  on which the
Faddeev-Popov determinant vanishes).

The Gribov's quantisation prescription is explicitly realised in the lattice
formulation.  Indeed, the procedure of Landau gauge fixing on the lattice
(see \cite{Giusti:2001xf} for a review)  constraints all the eigenvalues of
the Faddeev-Popov operator to be positive.  Thus lattice gauge
configurations in Landau gauge are located inside the Gribov region, and all
calculations on the lattice are done within the Gribov's quantisation
prescription. But we also know that this is not enough - the gauge is
not  uniquely fixed even in this case, and there are secondary solutions
(Gribov copies) inside the Gribov region
(\cite{Franke_Semenov_Tyan_Shanskii},\cite{Dell'Antonio:1989qg}). The
smaller region free of Gribov copies is called the fundamental modular
region 
(\cite{Franke_Semenov_Tyan_Shanskii},\cite{vanBaal:1991zw},
\cite{Zwanziger:1991ac}).
On a finite lattice the gauge may be fixed in a unique way, but that reveals
to be technically difficult (cf. the following section).

The quantities that are sensitive to the choice of the infrared quantisation
prescription are Green functions. A lot of work regarding the influence of
Gribov copies on lattice ghost and gluon  propagators in the case of $SU(2)$
and $SU(3)$ gauge groups has already been done 
(\cite{Cucchieri:1997dx},\cite{Bakeev:2003rr},\cite{Sternbeck:2004qk},%
\cite{Sternbeck:2004xr},\cite{Silva:2004bv},\cite{Sternbeck:2005tk},\cite{Sternbeck:2005vs}).
Summarising the results, the infrared divergence of the ghost propagator is
lessened when choosing  Gribov copies closer to the fundamental modular
region, and the gluon propagator remains  the same (within today's
statistical precision). In this paper we confirm these results in the case
of the $SU(2)$ gauge group, and present a study of the influence of Gribov
copies on the three-gluon vertex in symmetric and asymmetric kinematic
configurations. We also discuss in details the structure of minima of the
gauge-fixing  functional. The authors of \cite{Bakeev:2003rr} reported that
the number of minima of the gauge-fixing functional decrease with the
$\beta$ parameter (the bare lattice coupling). We perform a thorough
analysis of the volume dependence of this phenomenon. We define  a 
probability to find a lattice Gribov copy, and our computation shows that 
this probability increases considerably when the physical size of the
lattice exceeds  some critical value, around $2.75/\sqrt{\sigma}$, where
$\sigma$ is the string tension.

%
%
\section{Non-perturbative Landau gauge fixing on the lattice}
%
%

A lattice gauge configuration $U_{\mathcal{C}_0}$ generated during the simulation process is 
not gauge-fixed. One has to perform a gauge transformation $\{u(x)\}$ on it 
in order to move it along its gauge orbit up to the intersection with the surface $f_L\left[U_\mu(x)\right]=0$
specifying the Landau gauge-fixing condition. But there is no need to have an explicit 
form of $\{u(x)\}$. Instead we do an iterative minimisation process 
that starts at $U_{\mathcal{C}_0}$ and converges to the gauge fixed configuration $U^{(u)}_{\mathcal{C}}$
Let us first illustrate it on the example of the Landau gauge in the continuum limit.
For every gauge field $\mathcal{A}$ one defines a functional
\begin{equation}
\label{GaugeMinimisingFunctional}
F_\mathcal{A}[u(x)]=-\tr\int d^4x \mathcal{A}_{\mu}^{(u)}(x) \mathcal{A}_{\mu}^{(u)}(x)=
-\arrowvert\arrowvert\mathcal{A}^{(u)}\arrowvert\arrowvert^2,
\quad u(x)\in SU(N_c).
\end{equation}
Expanding it up to the second order around some group elements $\{u_0(x)\}$ we have (writing $u=e^{X}u_0$, $X\in\mathfrak{su}(N_c)$)
\begin{equation}
\label{GaugeMinimisingFunctionalEXP}
F_\mathcal{A}[u(x)]= F_\mathcal{A}[u_0(x)]+2\int d^4x \tr\left( X \partial_\mu \mathcal{A}^{(u_0)}_{\mu } \right)
+\int d^4 x \tr \left(X^\dagger \mathcal{M}_{\text{FP}}\left[\mathcal{A}^{(u_0)}\right]  X \right)+\ldots.
\end{equation}
Note that the quadratic form defining the second order derivatives is the Faddeev-Popov operator. 
Obviously, if $u_0$ is a local minimum we have a double condition
\begin{equation}
\left\{
	\begin{array}{l}
		\partial_\mu \mathcal{A}^{(u_0)}_{\mu } = 0 
		\\ 
		\mathcal{M}_{\text{FP}}\left[\mathcal{A}^{(u_0)}\right] \quad \text{is positive definite.}
	\end{array}
\right.
\end{equation}
Hence, the minimisation of the functional 
(\ref{GaugeMinimisingFunctional}) allows not only to fix the Landau gauge,
but also to obtain a gauge configuration inside the Gribov horizon (cf. Figure \ref{Gribov_horizon}).
%
\begin{figure}[ht]
\begin{center}
\epsfig{file=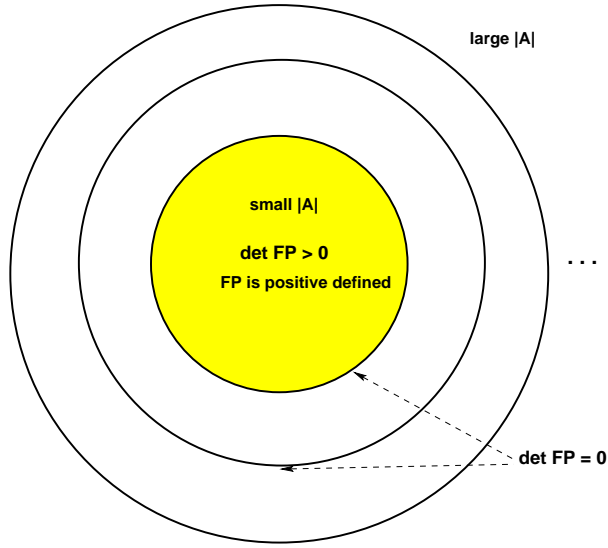,width=0.5\linewidth}  
\caption{\footnotesize Restriction of the integration domain in the partition function 
to the Gribov region (hatched), and the Gribov horizon. ``FP" is the Faddeev-Popov operator.}
\label{Gribov_horizon}
\end{center}
\end{figure}
%
On the lattice, the 
discretised functional (\ref{GaugeMinimisingFunctional}) reads
\begin{equation}
\label{LatticeGaugeMinimisingFunctional}
F_U[u]=-\frac{1}{V}\re\tr\sum_{x,\mu}u(x)U_\mu(x)u^{\dagger}(x+\emu),
\end{equation}
where $U_\mu(x)$ is the standard gauge link variable.
Then at a local minimum $u_0$ we have a discretised Landau gauge fixing condition 
\begin{equation}
\label{LatticeLandauGaugeCondition}
\sum_{\mu}
\left(
\mathcal{A}^{(u_0)}_{\mu}\left(x+\frac{\emu}{2}\right) -
\mathcal{A}^{(u_0)}_{\mu}\left(x-\frac{\emu}{2} \right)
\right)=0.
\end{equation}
that we write in a compact form $\nabla_\mu\mathcal{A}^{(u_0)}_\mu = 0$. Here
\begin{align}
\label{A_def}
\mathcal{A}_{\mu}\left(x+\frac{\emu}{2}\right) = \frac{U_{\mu}(x)-U_{\mu}^{\dagger}(x)}{2}.
\end{align}

The second derivative (the equivalent of the second-order term in
(\ref{GaugeMinimisingFunctionalEXP})) can  be written as
\begin{equation}
\frac{d^2}{ds^2}F_U\left[ u(s,x) \right]  = -\frac{1}{V} \left( \omega, \nabla_\mu \mathcal{A}^{\prime}_\mu \right)
\end{equation}
where
\begin{equation}
\label{A_PRIME}
\mathcal{A}^{\prime}_\mu = 
\frac{1}{2} 
\left( 
	-\omega(x) U^{(u)}_\mu(x) + U^{(u)}_\mu(x) \omega(x+\emu) + \omega(x+\emu) U^{(u)\dagger}_\mu(x) - 
	U^{(u)\dagger}_\mu(x) \omega(x)
\right).
\end{equation}
This defines a quadratic form $\left(  \omega, \MFPlat[U] \omega  \right)$ where the 
operator $\MFPlat[U]$ is the lattice version of the Faddeev-Popov operator (\cite{Zwanziger:1993dh}). 
It reads 
\begin{align}
\label{MFPlat}
\nonumber
\Big(\MFPlat[U] & \omega\Big)^{a} (x) = \frac{1}{V}\sum_{\mu}\biggl\{ S_{\mu}^{ab}(x)
          \left(\omega^{b}(x+\emu)-\omega^{b}(x)\right)
        - (x \leftrightarrow x-\emu) + \\
 & + \frac{1}{2}f^{abc}\left[
             \omega^{b}(x+\emu)A_{\mu}^{c}\left(x+\frac{\emu}{2}\right)
           - \omega^{b}(x-\emu)A_{\mu}^{c}\left(x-\frac{\emu}{2}\right) \right]
          \biggr\},
\end{align}
where
\begin{align}
& S_{\mu}^{ab}(x) = - \frac{1}{2}\tr\left(\left\{t^{a},t^{b}\right\} \left(U_{\mu}(x)+U_{\mu}^{\dagger}(x)\right)\right).
\end{align}
It is straightforward to check that in the continuum limit $a\rightarrow 0$
one finds the familiar expression of the Faddeev-Popov operator.

For the Landau gauge fixing in our numerical simulation we have used the
Overrelaxation algorithm with $\omega=1.72$. We stop the iteration process of
the minimising algorithm when the following triple condition is fullfield:
\begin{align}
\label{StopParameters}
\frac{1}{V (N_c^2 -1)}\sum_{x,\mu}
\tr \left[  
\left(
\nabla_\mu\mathcal{A}^{(u_0)}_\mu
\right)
\left( \nabla_\mu\mathcal{A}^{(u_0)}_\mu \right)^{\dagger}\right] 
& \leq \Theta_{\max_x \arrowvert \partial_\mu A^a_\mu\arrowvert} = 10^{-18}
\nonumber
\\
\frac{1}{V (N_c^2 -1)}
\left\arrowvert
\sum_x \tr\left[ u^{(n)}(x) - \id \right]
\right\arrowvert & \leq \Theta_{\delta u } = 10^{-9}
\nonumber
\\
\forall a,t_1,t_2 \quad 
\left \arrowvert 
\frac{ \text{A}^a_0 (t_1)-\text{A}^a_0 (t_2)  }{\text{A}^a_0 (t_1)+\text{A}^a_0 (t_2) }
\right \arrowvert
& \leq \Theta_{\delta \text{A}_0 } = 10^{-7}.
\end{align}
where $u^{(n)}(x)$ is the matrix of the gauge transformation $u(x)$ at the iteration step $n$, 
and the charge 
\begin{equation}
\text{A}^a_0(\vec{x},t) = \int d^3 \vec{x} A^a_0(\vec{x},t)
\end{equation}
must be independent of $t$ in Landau gauge when periodical boundary conditions for the gauge 
field are used. The choice of  numerical values for the stopping parameters 
is discussed at the end of the next section.

%
%
\section{The landscape of minima of the functional $F_U$}
%
%
\label{sec_b_critiq}

The functional $F_U$ (\ref{LatticeGaugeMinimisingFunctional}) has a form similar to 
the energy of a spin glass. The last is known to
posses a very large number of metastable states, i.e. spin configurations whose
energy increases when any spin is reversed. Typically, the number of these states grows exponentially
with the number of spins.
\begin{figure}[h]
\begin{center}
\epsfig{file=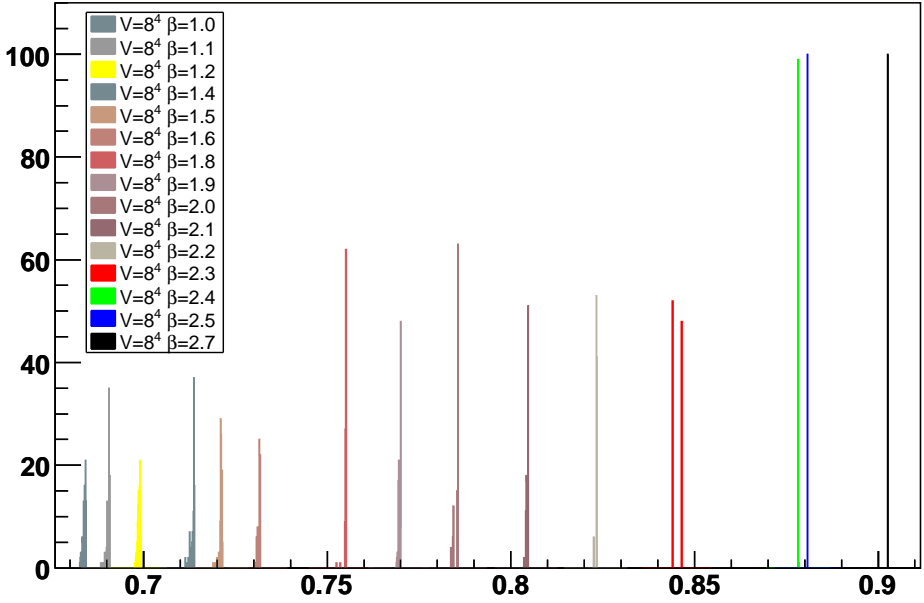,width=0.9\linewidth}  
\caption{\footnotesize Histogram of minima values $F_{\text{min}}$ for
different $\beta$ for \emph{one particular}  gauge-field configuration,
$\arrowvert \partial_\mu A^a_\mu(x) \arrowvert \le 10^{-9}$. Here
$N_\text{GF}=100$. The histograms are ranged in the order the increase of
the $\beta$ parameter.}
\label{Fmin_ditrib_beta_8}
\end{center}
\end{figure}

\begin{figure}[h]
\begin{center}
\epsfig{file=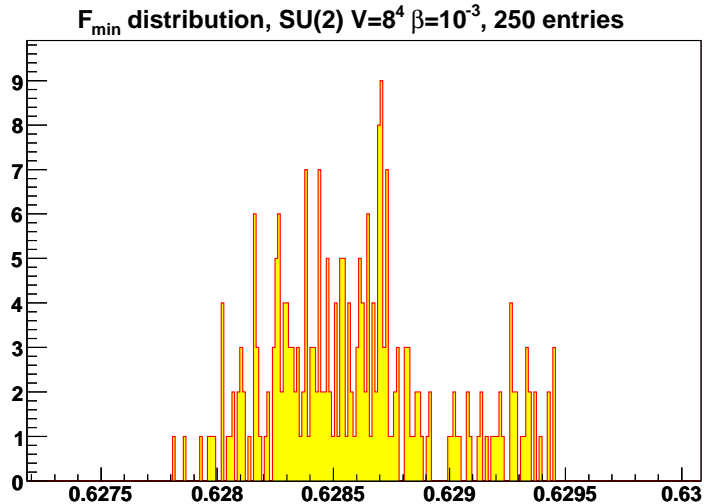,width=0.7\linewidth}  
\caption{\footnotesize Histogram of minima values $F_{\text{min}}$ for very a small value of $\beta=10^{-3}$,
$\arrowvert \partial_\mu A^a_\mu(x) \arrowvert \le 10^{-9}$.}
\label{Beta_zero}
\end{center}
\end{figure}
%
%
%
%
\begin{figure}[p]
\begin{center}
\epsfig{file=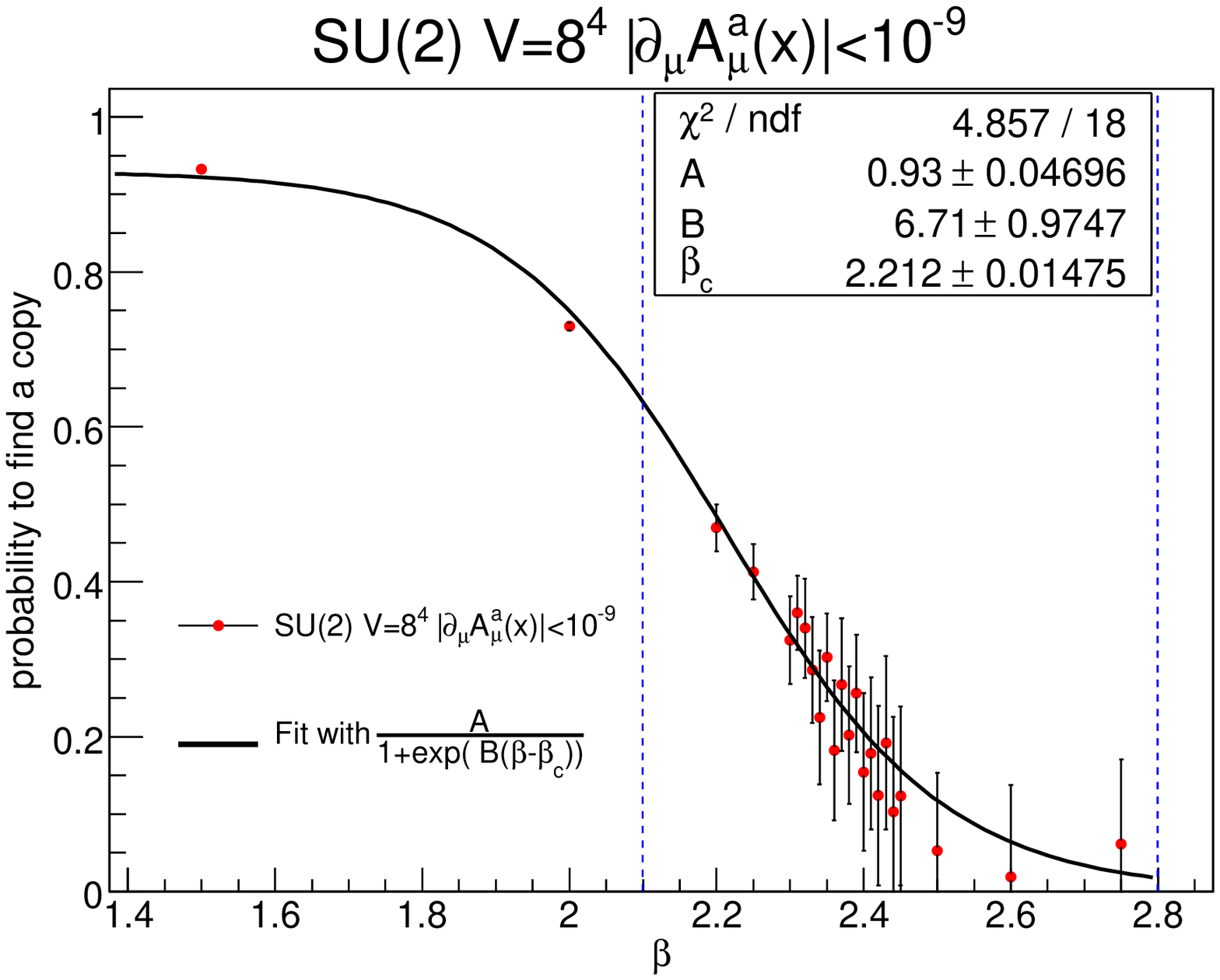,width=0.85\linewidth}  
\caption{\footnotesize Probability (averaged over gauge orbits) to find a secondary minimum as a function of $\beta$ at volume $V=8^4$.}
\label{PROBA_GRIBOV_8}
\end{center}
\end{figure}
%
%
%
\begin{figure}[p]
\begin{center}
\epsfig{file=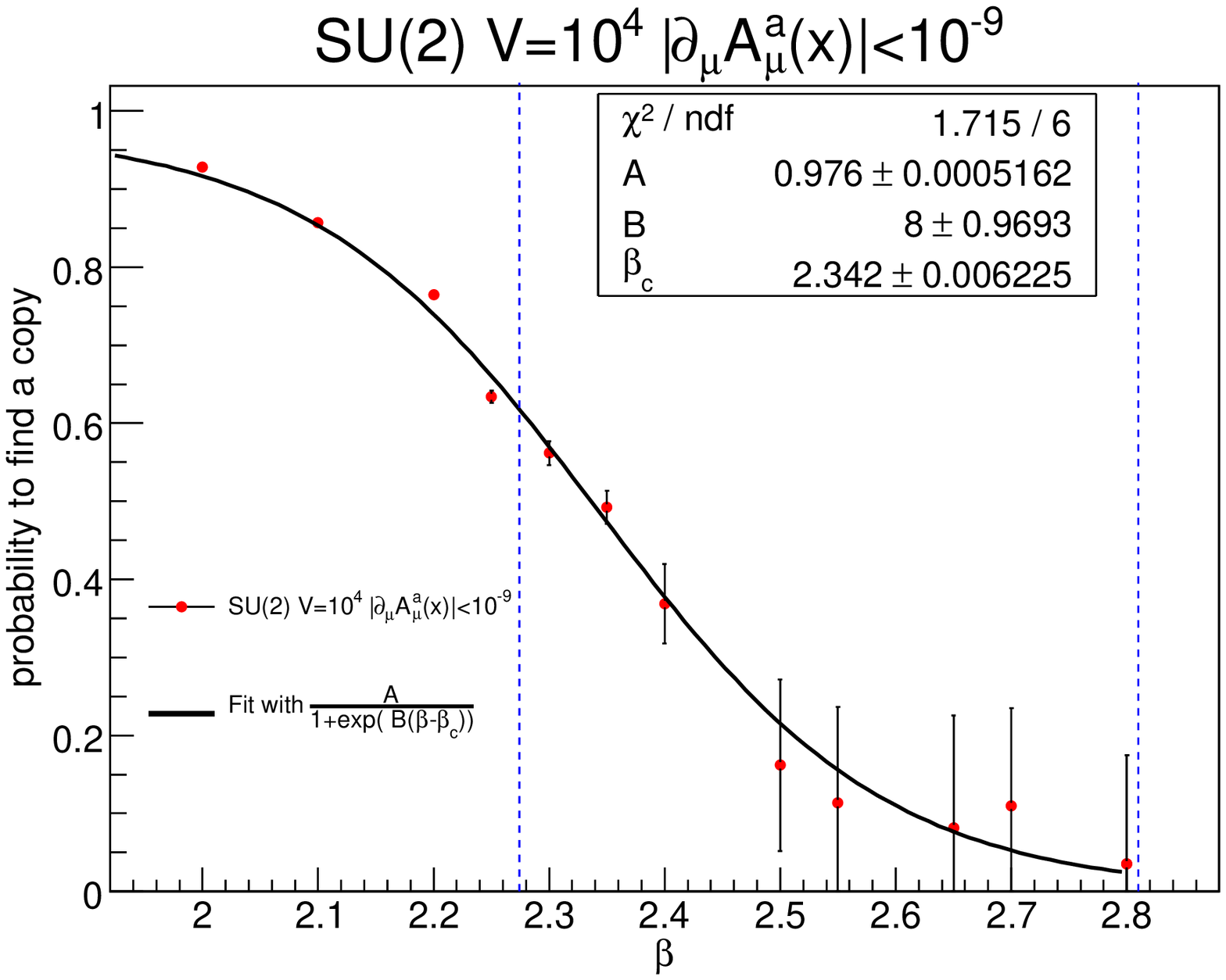,width=0.85\linewidth}  
\caption{\footnotesize Probability (averaged over gauge orbits) to find a secondary minimum as a function of $\beta$ at volume $V=10^4$.}
\label{PROBA_GRIBOV_10}
\end{center}
\end{figure}
%
%
%
\begin{figure}[p]
\begin{center}
\epsfig{file=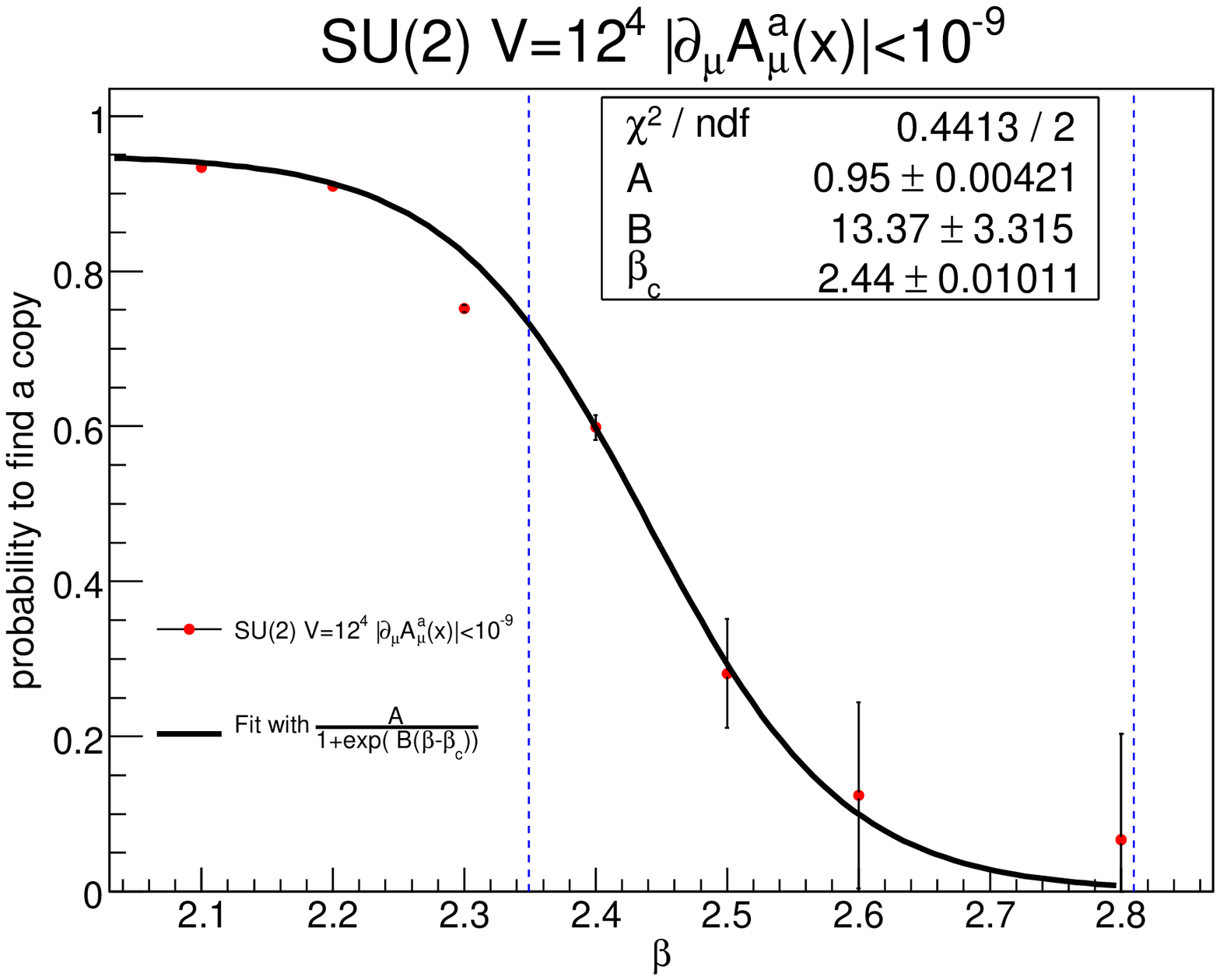,width=0.85\linewidth}  
\caption{\footnotesize Probability (averaged over gauge orbits) to find a secondary minimum as a function of $\beta$ at volume $V=12^4$.}
\label{PROBA_GRIBOV_12}
\end{center}
\end{figure}
%
%
%
\begin{figure}[p]
\begin{center}
\epsfig{file=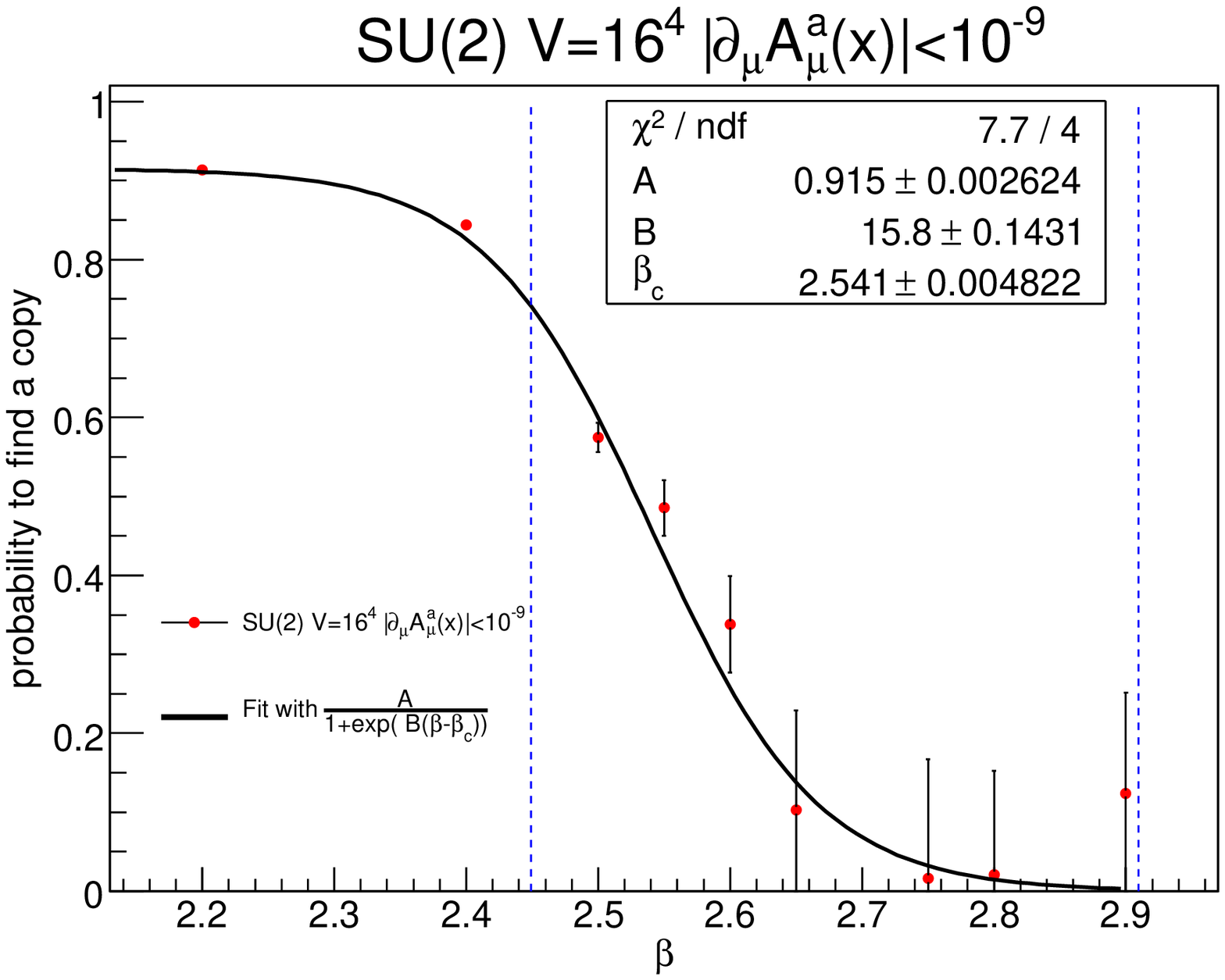,width=0.85\linewidth}  
\caption{\footnotesize Probability (averaged over gauge orbits) to find a secondary minimum as a function of $\beta$ at volume $V=16^4$.}
\label{PROBA_GRIBOV_16}
\end{center}
\end{figure}
%
%

Let us consider the landscape of the functional $F_U$. One of its
characteristics is  the distribution of values at minima $F_{\text{min}}$ of
$F_U$. We know that for small magnitudes  of the gauge field all the link
matrices $U_\mu(x)$ (they play the role of couplings between the ``spin" 
variables) are close to the identity matrix, and thus the minimum is
unique.  Their number increases when the bare lattice coupling $\beta$
decreases, because the typical  magnitude of the phase of $U_\mu(x)$ grows
in this case and thus link  matrices move farther from the identity matrix.
The number of minima also increases with the number of links (at fixed
$\beta$) because in this case there are more degrees of freedom in the
system.   At Figure \ref{Fmin_ditrib_beta_8} we present typical  histograms of
the distribution of $F_\text{min}$ for one  given gauge configuration, as a
function of the $\beta$ parameter.  We see that when $\beta$ is large
($\beta \gtrsim 2.5$ for the volume $V=8^4$ considered at Figure
\ref{Fmin_ditrib_beta_8}), we typically find only one value of
$F_\text{min}$. On the contrary, for very small values of $\beta$  we find
many different minima (cf. Figure \ref{Beta_zero}).  These values of  $\beta$
correspond to almost random links $U_\mu(x)$.  In the following section we
show that field configurations having  the same $F_\text{min}$ are in fact
equivalent up to a global gauge transformation.

The probability to find a secondary minimum depends on the value of the $\beta$ parameter.
We can calculate this probability  in the following way. For each orbit
we fix the gauge $N_\text{GF}$ times, each gauge fixing starts after a random gauge transformation 
of the initial field configuration. We thus obtain a distribution of minima $F_\text{min}$.
This distribution gives us the number of minima $N(F^{i})$ as a function of 
the value of $F^{i} \equiv F^{i}_{\text{min}} $. The relative frequency of a minimum $F^i$ is defined by 
\begin{equation}
\omega_i = \frac{N(F^{i})}{\sum_{i} N(F^{i}) },
\end{equation}
where $\sum_{i} N(F^{i})  = N_\text{GF}$. Then the weighted mean number of copies per value of $F_\text{min}$ is 
given by
\begin{equation}
\overline N = \sum_i \omega_i N(F^{i}).
\end{equation}
This allows us to define a probability to find a secondary minimum when
fixing the Landau gauge for a gauge field configuration~:
\begin{equation}
p_{\text{1conf}} = 1 - \frac{\overline N}{\sum_{i} N(F^{i}) }.
\end{equation}
If one finds the same value of $F_\text{min}$ for all $N_{\text{GF}}$ tries then this probability is zero.
On the contrary, if all $F_i$ are different then $p$ is close to one.

Having the probability to find a secondary minimum when fixing the gauge for
\emph{one particular} configuration we can calculate the Monte-Carlo
\emph{average} $\left\langle \star \right\rangle$ on gauge orbits, i.e. on 
``spin couplings'' $U_\mu(x)$. We obtain finally the overall probability
to find a secondary minimum during a numerical simulation: 
\begin{equation}
P(\beta)\equiv  \left\langle p_{\text{1conf}} \right\rangle = 1 - \left\langle \frac{\overline N}{\sum_{i} N(F^{i}) } \right\rangle.
\end{equation}

We have performed simulations in the case of $SU(2)$ latice gauge theory at
volumes $V=\{8^4, 10^4, 12^4, 16^4\}$  for $\beta$ varying from $1.4$ to
$2.9$. For each value of $\beta$ we generated $100$ independent Monte-Carlo 
gauge configurations, and we fixed the gauge $N_\text{GF} = 100$ times for
every configuration. Between each gauge fixing a random gauge transformation
of the initial gauge configuration was performed, and the minimising
algorithm stops  when the triple condition (\ref{StopParameters}) is
satisfied. The resulting probability to find  a secondary minimum is
presented at (Figures \ref{PROBA_GRIBOV_8}-\ref{PROBA_GRIBOV_16}).

As expected, the probability is small when $\beta$ is large, and it is close
to one when $\beta$ is small. The dispersion was calculated using the
Jackknife method. The physical meaning of this dispersion is the following:
when the error is small, all gauge configurations have a similar number of
secondary minima. On the contrary, this dispersion is large if there are
some exceptional gauge configurations having a different number of copies.
At small $\beta$ almost all gauge configurations have many secondary minima,
that is why the dispersion of the probability is small. At large $\beta$
almost all gauge configurations  have a unique minimum, but some of them can
have copies. This may considerably increase the dispersion of the
probability. The appearance  of exceptional gauge configuration possessing a
large density of close-to-zero eigenvalues of the Faddeev-Popov operator has
been recently reported~\cite{Sternbeck:2005vs}. Probably these fields are
related  with those having a  lot of secondary minima at large $\beta$'s,
and this correlation deserves a separate study.

We can fit the data from Figures \ref{PROBA_GRIBOV_8}-\ref{PROBA_GRIBOV_16} with an
empirical formula
\begin{equation}
P(\beta) = \frac{A}{1+e^{B(\beta - \beta_c)}}
\end{equation}
in order to define a characteristic coupling $\beta_c$ when the probability
to find a copy decreases considerably. One can define $\beta_c$ as
corresponding to the semi-heights of the probability function $P(\beta)$. At
this value of $\beta$ an {\it equally probable} secondary attractor of the
functional $F_U$ appears. The fit has been performed for the points between
dashed lines at Figures \ref{PROBA_GRIBOV_8}-\ref{PROBA_GRIBOV_16}, and the
results for the fit parameters are given ibidem and in Table \ref{lambda_c}.
We see that $\beta_c$ depends  on the volume of the lattice. Let us check
whether these values correspond to some physical scale. According to works
\cite{Bloch:2003sk},\cite{Fingberg:1992ju} one has the following expression
for the string tension $\sigma$ for $\beta\ge 2.3$:
\begin{equation}
\label{Spacing}
[\sigma a^2] (\beta) \simeq
 e^{-\frac{4\pi^2}{\beta_0}\beta + \frac{2\beta_1}{\beta_0^2}\log{\left( \frac{4\pi^2}{\beta_0}\beta \right)} + \frac{4\pi^2}{\beta_0}\frac{d}{\beta} + c}
\end{equation}
with $c= 4.38(9)$ and $d=1.66(4)$. Using this formula,  we define a characteristic scale 
corresponding to the critical values $\beta_c$ from (Figures \ref{PROBA_GRIBOV_8}-\ref{PROBA_GRIBOV_16}) :
$$
\lambda_c = a(\beta_c)\cdot L
$$
in the string tension units, $L$ is the length of the lattice. In the last column of Table \ref{lambda_c} we summarise the results.
\begin{table}[h]
\centering
\begin{tabular}{l|l|c|c}
$L$           &  $\beta_c$  & $\chi^2/\text{ndf}$ & $\lambda_c$, in units of $1/\sqrt{\sigma}$ 
\\ \hline 
$8$           &  $2.221(14)$ & $0.27$ & $3.85(31)$ 
\\ \cline{4-4}
$10$          & $2.342(6)$   & $0.28$ & $3.20(21)$ 
\\ 
$12$          & $2.44(1)$    & $0.22$ & $2.78(20)$
\\ 
$16$          & $2.541(5)$   & $1.92$ & $2.68(17) $
\\ 
\end{tabular}
\caption{\footnotesize The characteristic length defining the appearance of
secondary minima. The errors for $\lambda_c$  include errors for $d$ and $c$
parameters, and the fitted error for $\beta_c$}
\label{lambda_c}
\end{table}
We see that for the values of $\beta_c$ in the scaling regime when the
formula (\ref{Spacing}) is applicable ($\beta \ge 2.3$) we obtain compatible
values for the physical length $\lambda_c$. 

This suggests that lattice Gribov copies appear when the physical size of
the lattice exceeds  a critical value of around $2.75/\sqrt{\sigma}$. At
fist approximation $\lambda_c$ is scale  invariant, but a slight dependence
in the lattice spacing remains.  

In principle, the parameter $\beta_c$ can be calculated with good precision.
One should do it in the case of $SU(3)$ gauge  group, because the dependence
of the lattice spacing on the bare coupling is softer, and the scaling of
the theory  has been better studied than in the case of the $SU(2)$ theory.

Let us discuss the dependence of above results on the choice of the stopping
parameters (\ref{StopParameters}). In the Table \ref{TableStopPatameters} we
present the probabilities $P(2.0),P(2.3),P(2.8)$, calculated on a $8^4$
lattice with different values of the parameters (\ref{StopParameters}). One
can see that our choice of stopping parameters in (\ref{StopParameters}) is
strict enough to ensure the  independence of our results on the further
increase of the parameter $\Theta_{\max_x \arrowvert \partial_\mu
A^a_\mu\arrowvert}$.

%
%
\section{Green functions and the lattice Gribov copies}
%
%

%
\subsection{Two- and three-point lattice Green functions}
%

Lattice Green functions are computed as Monte-Carlo averages over
gauge-fixed gluon  field configurations.

Gluonic Green functions are defined using the definition (\ref{A_def}) of
the gauge field. The ghost propagator $F^{(2)ab}(x-y)$ is computed, with the
algorithm given in~\cite{Boucaud:2005gg}, by numerical inversion  of the
Faddeev-Popov operator~(\ref{MFPlat}), e.g.
\begin{align}
\label{eq:Ghost}
F^{(2)}(x-y)\delta^{ab} \equiv \left\langle\left({\MFPlat}^{-1}[A]\right)^{ab}_{xy}\right\rangle.
\end{align}
\\
The gluon propagator in Landau gauge may be parametrised as
\begin{equation}
G^{(2)ab}_{\mu\nu}(p,-p) \equiv
\left\langle
\widetilde{A}^a_\mu(-p)  \widetilde{A}^b_\nu(p)
\right\rangle
= 
\delta^{ab}\left(  \delta_{\mu\nu} - \frac{p_\mu p_\nu}{p^2}  \right)
G^{(2)}(p^2)
\end{equation}
completed with
\begin{equation}
G^{(2)ab}_{\mu\nu}(0,0) = 
\delta^{ab}  \delta_{\mu\nu} G^{(2)}(0).
\end{equation}
\\
The ghost propagator is parametrised in the usual way:
\begin{equation}
\widetilde{F}^{(2)ab}(p,-p) \equiv
\left\langle
\widetilde{c}^a (-p)  \widetilde{\overline{c}}^b(p) \right\rangle 
=
\delta^{ab} F^{(2)}(p^2)
\end{equation}

The coupling constant can be defined non-perturbatively by the amputation of
a  three-point Green-functions from its external propagators. But this
requires to fix the  kinematic configuration of the three-point
Green-function at the normalisation  point. On the lattice one usually uses
either a fully symmetric kinematic configuration (denoted $\text{MOM}$)  or
a zero point kinematic configuration with one vanishing external momentum
(denoted generically  $\widetilde{\text{MOM}}$). In what follows we only
speak about gluonic three-point functions.
\subsubsection*{Symmetric case} 
There are only two independent tensors in Landau gauge in the case of the
symmetric three-gluon Green function~\cite{Boucaud:1998bq}:
\begin{align}
&\mathcal{T}^{[1]}_{\mu_{1},\mu_{2},\mu_{3}}(p_1,p_2,p_3) = 
\delta_{\mu_{1}\mu_{2}}(p_1-p_2)_{\mu_{3}} + 
\delta_{\mu_{2}\mu_{3}}(p_2-p_3)_{\mu_{1}} + 
\delta_{\mu_{3}\mu_{1}}(p_3-p_1)_{\mu_{2}}
\\&
\mathcal{T}^{[2]}_{\mu_{1},\mu_{2},\mu_{3}}(p_1,p_2,p_3) = 
\frac{(p_1 - p_2)_{\mu_{3}}  (p_2 - p_3)_{\mu_{1}}  (p_3 - p_1)_{\mu_{2}}  }{p^2}.
\end{align}
Then the three-gluon Green function in the $\text{MOM}$ scheme
($p_1^2=p_2^2=p_3^2=\mu^2$) can be parametrised as
\begin{equation}
\begin{split}
\left\langle 
\widetilde{A}^a_{\mu_{1}}(p_1) \widetilde{A}^b_{\mu_{2}}(p_2) \widetilde{A}^c_{\mu_{3}}(p_3)
\right\rangle  =
f^{abc} &
\Big[ 
G^{(3)\text{sym}}(\mu^2) \mathcal{T}^{[1]}_{\mu^{\prime}_{1},\mu^{\prime}_{2},\mu^{\prime}_{3}}(p_1,p_2,p_3)
\prod_{i=1,3} 
	\left( 
		\delta_{\mu^{\prime}_{i} \mu_{i}} - \frac{{p_i}_{\mu^{\prime}_{i}} {p_i}_{\mu_{i}} }{\mu^2}
	\right) +
\\ & + H^{(3)}(\mu^2) \mathcal{T}^{[2]}_{\mu_{1},\mu_{2},\mu_{3}}(p_1,p_2,p_3)
\Big]
\end{split}
\end{equation}
The scalar function $G^{(3)\text{sym}}(\mu^2)$, proportional to the coupling
$g_0$ at tree level, may be extracted by the following projection:
\begin{equation}
\begin{split}
G^{(3)\text{sym}}(\mu^2) = &
\left( 
\mathcal{T}^{[1]}_{\mu^{\prime}_{1},\mu^{\prime}_{2},\mu^{\prime}_{3}}(p_1,p_2,p_3)
\prod_{i=1,3} \left( 
\delta_{\mu^{\prime}_{i} \mu_{i}} - \frac{{p_i}_{\mu^{\prime}_{i}} {p_i}_{\mu_{i}} }{\mu^2}\right)
+
\frac{1}{2}\mathcal{T}^{[2]}_{\mu^{\prime}_{1},\mu^{\prime}_{2},\mu^{\prime}_{3}}(p_1,p_2,p_3)
\right)\times
\\ & \times
\frac{1}{18\mu^2}
\frac{f^{abc}}{N_c(N_c^2-1)} 
\left\langle 
	\widetilde{A}^a_{\mu_{1}}(p_1) \widetilde{A}^b_{\mu_{2}}(p_2)
\widetilde{A}^c_{\mu_{3}}(p_3)
\right\rangle.
\end{split}
\end{equation}
\subsubsection*{Asymmetric case} The three-gluon Green function with one
vanishing  external propagator (\cite{Alles:1996ka},\cite{Boucaud:1998bq}) can
be parametrised as
\begin{equation}
G^{(3)abc}_{\mu\nu\rho}(p,0,-p) \equiv
\left\langle
\widetilde{A}^a_\mu(-p)  \widetilde{A}^b_\nu(p) \widetilde{A}^c_\rho(0)
\right\rangle
=2f^{abc} p_\rho \left(  \delta_{\mu\nu} - \frac{p_\mu p_\nu}{p^2}  \right)
G^{(3)\text{asym}}(p^2),
\end{equation}
and thus
\begin{equation}
G^{(3)\text{asym}}(p^2) =
\frac{1}{6p^2}\frac{f^{abc}}{N_c(N_c^2-1)}\delta_{\mu\nu}p_\rho
G^{(3)abc}_{\mu\nu\rho}(p,0,-p).
\end{equation}

\subsubsection*{Gauge coupling} Using the scalar functions $G^{(2)}$ and
$G^{(3)}$ (the last stands for $G^{(3)\text{asym}}$  or $G^{(3)\text{sym}}$),
the gauge coupling at the renormalisation scale $\mu^2$ is defined by
\begin{equation}
g_R(\mu^2) = \frac{G^{(3)}(p_1^2,p_2^2,p_3^2)}{G^{(2)}(p_1^2) G^{(2)}(p_2^2)
G^{(2)}(p_3^2)} Z_3^{3/2}(\mu^2)
\end{equation}
in the case of three-gluon vertices, where the choice of $p_i$ determines
the renormalisation scheme  ($\widetilde{\text{MOM}}$ or $\text{MOM}$).
In paper~\cite{Boucaud:2002fx} it is shown that in the  $\text{MOM}$ schemes
$g_R(\mu^2)$ vanishes in the zero momentum limit. It is interesting to
investigate the possible influence of Gribov copies on this behaviour. We
address this question in the following section. 
%
\subsection{Influence of the lattice gauge fixing}
%

A natural question that arises after the study of the distribution of
$F_\text{min}$ is whether the gauge configurations having the same value of
$F_\text{min}$ are equivalent, i.e. they differ only by a global gauge
transformation. %
\begin{table}[h]
\centering
\begin{footnotesize}
\begin{tabular}{r||c|c||c}
\hline
$F_\text{min}\text{~~~~~~~~}$  & $-0.871010810260$ & $-0.871010810260$ & $-0.870645877060$ 
\\ \hline 
$V\cdot G^{(2)}(p^2):\quad p^2=0$ & $4249297~$ & $4249295~$ & $3322788~$
\\ 
$p^2=1$ & $2012518~$ & $2012516~$ & $2006186~$
\\ 
$p^2=2$ & $1215762~$ & $1215762~$ & $1362671~$
\\ 
$p^2=3$ & $834876.3$ & $834876.4$ & $798032~~$
\\ 
$p^2=4_{~p^{[4]}=16}$ & $620065.2$ & $620067.3$ & $434235.6$
\\ 
$p^2=4_{~p^{[4]}=4~}$ & $521585.0$ & $521585.2$ & $556509.6$
\\ 
$p^2=5$ & $410698.8$ & $410698.5$ & $440623.9$
\\ \hline
\end{tabular}
\caption{\footnotesize Gluon correlator calculated on different gauge
configurations having the same value of $F_\text{min}$ (columns 2 and 3),
compared to values from the "secondary" minimum (column 4). We present data for
all $H_4$ group orbits for the momentum $p^2=4$
(\cite{Becirevic:1999hj},\cite{Boucaud:2005gg}). Notice that in all this paper 
the momenta are given in units of $2\pi/(aL)$. Simulation has been performed
with parameters $V=16^4,\beta=2.4$}  \label{same_Fmin_gluon_propagator}
\vspace*{0.5cm}
\begin{tabular}{r||c|c||c}
\hline
$F_\text{min}\text{~~~~~~~~}$  & $-0.871010810260$ & $-0.871010810260$ & $-0.870645877060$ 
\\ \hline 
$F^{(2)}(p^2)\quad p^2=1$ & $14.06473$ & $14.06473$ & $14.82984$
\\ 
$p^2=2$ & $6.278253$ & $6.278253$ & $6.736338$
\\ 
$p^2=3$ & $3.757531$ & $3.757531$ & $3.939130$
\\ 
$p^2=4_{~p^{[4]}=16}$ & $2.929602$ & $2.929602$ & $2.705556$
\\ 
$p^2=4_{~p^{[4]}=4~}$ & $2.599088$ & $2.599088$ & $2.775566$
\\ 
$p^2=5$ & $2.071200$ & $2.071200$ & $2.011100$
\\ \hline
\end{tabular}
\caption{\footnotesize Ghost correlator calculated on different gauge configurations
having the same value of $F_\text{min}$ (columns 2 and 3), compared to
values from the "secondary" minimum (column 4). We present data for all
$H_4$ group orbits for the momentum $p^2=4$ (\cite{Becirevic:1999hj},\cite{Boucaud:2005gg}).
Simulation has been performed with
parameters $V=16^4,\beta=2.4$} 
\label{same_Fmin_ghost_propagator}
\end{footnotesize}
\end{table}

This can be checked by calculating the two-point gluonic correlation function on the gauge
configuration . Indeed, according to the lattice
definition of the gauge field that we used (\ref{A_def}),
\begin{equation}
G^{(2)}_{\text{1 conf}}(x-y) \propto
\tr  \left[ \left(U_\mu(x) - U_\mu^\dagger(x)\right) \cdot \left(U_\nu(y) -
 U_\nu^\dagger(y) \right) \right].
\end{equation}
Applying a global gauge transformation $U_\mu(x)\rightarrow V U_\mu(x)
V^\dagger$ we see that the  gluon propagator remains unchanged. This is also
the case of the ghost propagator scalar function. We have checked
numerically that the values of the gluon and the ghost propagators in
Fourier space  are the same for gauge configurations having the same
$F_\text{min}$  (cf.
Tables \ref{same_Fmin_gluon_propagator},\ref{same_Fmin_ghost_propagator} ). For
comparison, the data from a lattice Gribov copy (the same gauge
configuration but having different value of $F_\text{min}$) is also given.
We see that the values are almost the same for gauge configurations having
the same $F_\text{min}$  contrarily to those coming from different minima.
In fact, taking in account rounding errors appearing during the calculation,
we may say that gauge configurations having the same $F_\text{min}$
\emph{are} equivalent.

The second question that has already been considered by (
\cite{Cucchieri:1997dx},\cite{Bakeev:2003rr},\cite{Sternbeck:2004qk},%
\cite{Sternbeck:2004xr},\cite{Silva:2004bv},\cite{Sternbeck:2005tk},\cite{Sternbeck:2005vs}
) is the dependence of Green functions on the choice of the minimum. For
this purpose we have performed the following simulation: for every of the
$100$ gauge configurations used to compute Green functions the gauge was
fixed $100$ times, and the Monte-Carlo average  was computed with respect to
the ``first copy'' (fc) found by the minimisation algorithm and the ``best
copy'' (bc), having the smallest value of $F_\text{min}$. We have calculated
the gluon and  the ghost propagators, and also the three-gluon Green
functions in symmetric and  asymmetric kinematic configurations. The
simulations have been performed on lattices of volumes  $8^4$ and $16^4$ for
$\beta=2.1, 2.2, 2.3$. At these values of $\beta$ we are sure to have 
lattice Gribov copies.  The results are given in Tables
\ref{VOL_8_BETA_21_GHOST_PROPAGATOR}-\ref{VOL_16_BETA_23_GLUON_VERTEX_SYM}.
All data is given in lattice units. We present data for all $H_4$ group orbits 
(\cite{Becirevic:1999hj},\cite{Boucaud:2005gg}) for considered momenta, 
and we have checked that our results for two-point Green functions are consistent 
with numerical values given in~\cite{Cucchieri:1997dx},\cite{Bakeev:2003rr}. 
We conclude from Tables \ref{VOL_8_BETA_21_GHOST_PROPAGATOR}-\ref{VOL_16_BETA_23_GLUON_VERTEX_SYM}
that the ghost propagator is quite sensitive to the choice of the minimum - in the
case of (bc) the infrared divergence is lessened. No systematic effect could
be found for gluonic two- and three-point Green functions, the values in  
the cases of (fc) and (bc) being compatible within the statistical errors.
\begin{table}
\centering
\begin{footnotesize}
\begin{tabular}{ccccc}
\hline
$\beta$ & $L$ & $p^2$ & $F^{(2)}_{\text{fc}}(p^2)-F^{(2)}_{\text{bc}}(p^2)$ & $\frac{F^{(2)}_{\text{fc}}(p^2)-F^{(2)}_{\text{bc}}(p^2)}{F^{(2)}_{\text{bc}}}$
\\ \hline
$2.1$ & $8~$ &$1$ & $0.211$ & $0.045$ 
\\ 
$2.1$ & $16$ &$4_{~p^{[4]}=16}$ & $0.145$ & $0.033$ 
\\ \hline
$2.2$ & $8~$ &$1$ & $0.078$ & $0.019$ 
\\ 
$2.2$ & $16$ &$4_{~p^{[4]}=16}$ & $0.023$ & $0.006$ 
\\ \hline
$2.3$ & $8~$ &$1$ & $0.086$ & $0.024$ 
\\ 
$2.3$ & $16$ &$4_{~p^{[4]}=16}$ & $0.114$ & $0.034$ 
\\ \hline
\end{tabular}
\caption{\footnotesize Volume dependence of the ghost propagators, from Tables ~\ref{VOL_8_BETA_21_GHOST_PROPAGATOR}-
\ref{VOL_16_BETA_23_GHOST_PROPAGATOR} }
\label{bc_fc}
\end{footnotesize}
\end{table}

It is conjectured in~\cite{Zwanziger:2003cf} that in the infinite volume limit
the  expectation values calculated by integration over the Gribov region and
the fundamental  modular region become equal. However, the ghost propagator
depends on this choice, even for  volumes larger than the critical volume
defined in section \ref{sec_b_critiq}, see 
Table~\ref{VOL_16_BETA_21_GHOST_PROPAGATOR} where there is a four $\sigma$
discrepancy for $p^2 = 1$.  This dependence has been found to decrease slowly
with the volume \cite{Sternbeck:2005tk}.  The results of
section~\ref{sec_b_critiq} indicate that the convergence can only happen beyond
the critical size. To check this we compare the fc-bc values of the ghost
propagator, at one physical value of the momentum, for the orbit $p^2=1$ on
a $8^4$ lattice and the orbit $p^2=4, p^{[4]}=16$ on the $16^4$
lattice~\footnote{Remember that the momentum in physical units is equal to 
$2\pi\,p /(La)$ in our notations.} at the same $\beta$ (see Table~\ref{bc_fc}).
It happens indeed that the decrease is observed only at $\beta=2.1$ and $2.2$,
in accordance with Table~\ref{lambda_c}. However, these values of $\beta$ are
not in the scaling regime, and thus a study on larger lattices would be
welcome. It is not surprising that the ghost propagator depends on the bc/fc
choice: the bc corresponds to the  fields further from the Gribov horizon where
the Faddeev-Popov operator has a zero mode, whence the  inverse Faddeev-Popov
operator (ghost propagator) is expected to be smaller as observed.  The
correlation between the bc/fc choice and the gluon propagator is not so direct.
\begin{table}
\centering
\begin{footnotesize}
\begin{tabular}{cc|cc}
\hline \hline \\
$\beta$ & $L$ & $\overline{\langle F_\text{min} \rangle}_{\{U\}}$ & $\delta\overline{\langle F_\text{min} \rangle}_{\{U\}}$
\\ \hline
$2.2$ & $8~$ & $-0.8236$ & $0.003744$ 
\\ 
      & $10$ & $-0.8262$ & $0.002367$ 
\\ 
      & $12$ & $-0.8272$ & $0.001377$ 
\\ 
      & $16$ & $-0.8279$ & $0.000802$ 
\\ \hline
$2.4$ & $8~$ & $-0.8642$ & $0.005270$ 
\\ 
      & $12$ & $-0.8669$ & $0.002739$ 
\\ 
      & $12$ & $-0.8686$ & $0.001849$ 
\\ 
      & $16$ & $-0.8702$ & $0.001003$ 
\\ \hline
\end{tabular}
\caption{\footnotesize Volume dependence of the Monte-Carlo+gauge orbit mean value at minima $F_\text{min}$ and the dispersion of this mean.} 
\label{Fmin_mean_volume_dep}
\end{footnotesize}
\end{table}
Another quantity is obviously strongly correlated to the bc/fc choice: 
the value of $F_\text{min}$. We tested the volume dependence of the 
Monte-Carlo+gauge orbit mean value of the quantity $F_\text{min}$  (see
Tab.\ref{Fmin_mean_volume_dep}). According to the argument given
in~\cite{Zwanziger:2003cf}, all minima  become degenerate in the infinite
volume limit, and closer to the absolute minimum (in the fundamental modular
region).  We see from the Table \ref{Fmin_mean_volume_dep} that their 
average value and dispersion decrease with  the volume at fixed $\beta$, 
in agreement with~\cite{Zwanziger:2003cf}. 

%
%
\section{Conclusions}
%
%

Our study showed that 
\begin{itemize}
\item Lattice Gribov copies appear and their number grows very fast when the
 physical  size of the lattice exceeds some critical value  
 $\approx 2.75/\sqrt{\sigma}$. This result is fairly independent of the 
 lattice spacing.

\item The configurations lying on the same gauge orbit and having the same
$F_\text{min}$ are equivalent, up to a global gauge transformation, and yield the same Green functions.
Those corresponding to minima of $F_{U}$ with different values of $F_\text{min}$ differ 
by a non-trivial gauge transformation, and thus they are not equivalent.

\item We confirm the result
(\cite{Cucchieri:1997dx},\cite{Bakeev:2003rr},\cite{Sternbeck:2004qk},%
\cite{Sternbeck:2004xr},\cite{Silva:2004bv},\cite{Sternbeck:2005tk},\cite{Sternbeck:2005vs})
that the divergence of the ghost propagator is lessened when choosing the
``best copy" (corresponding to the choice of the gauge configuration having
the  smallest value of $F_{U}$). We also showed that gluonic Green functions
calculated in the  ``first copy'' and ``best copy'' schemes are compatible
within the statistical error, no systematic  effect was found. We conclude
that gauge couplings that are defined by amputating the three-gluon 
vertices only slightly depend on the choice of the minimum of $F_U$. It
implies that the infrared behaviour of $g_R(\mu^2)$ reported 
in~\cite{Boucaud:2002fx} is not significantly influenced by lattice
 Gribov copies.

\item We found that the influence of Gribov copies on the ghost propagator
decreases with the volume when the physical lattice size is larger  than the
critical length discussed above. We also show that  the quantity $F_\text{min}$
decreases when the volume increases. These two points are in agreement with the
argument on the equality of the averages over the Gribov's region and the
fundamental modular region~\cite{Zwanziger:2003cf}.

\end{itemize}
The fact that the abundance of lattice Gribov copies is mainly  an
increasing function of the physical size of the lattice is  not too
surprising. The existence of Gribov copies is a non-perturbative phenomenon
and as such is related~\cite{Gribov:1977wm} to the infrared properties of
the Faddeev-Popov operator and to the confinement scale. It should then
dominantly depend on the infrared cut-off, the size (or volume) of the
lattice in physical units. A milder dependence on the ultraviolet cut-off,
the lattice spacing, is also expected but the limited accuracy concerning
the lattice spacing did not allow us to identify it.  In a recent
paper~\cite{Sternbeck:2005vs} the lowest eigenvalues of the Faddeev-Popov
operator have been computed. A dependence on the size of  the lattice is
seen. A detailed study of this dependence in connection  with our findings
in this paper would be useful. 

More generally, the question of extracting from lattice simulations
informations about Gribov copies in the continuum limit is not a simple
issue.  However, bulk quantities, such as the one we propose in this paper,
may be traced down to the continuum limit and provide precious information
about continuum Gribov copies. The total number of lattice Gribov copies in
a given gauge orbit may be large. All the minima of the functional
$F_\mathcal{A}[u(x)]$~(\ref{GaugeMinimisingFunctional})  are possible
end-points of the Landau gauge fixing algorithm, but their probability to be
selected  by the algorithm depends on the size of their domain of attraction
and might vary significantly. This probability is therefore expected to depend
on the attractor pattern and not on the choice of algorithm, which should be
checked. 

The explosion of the number of Gribov copies at larger volume does not 
contradict the statement that they a have decreasing influence on expectation
values~\cite{Zwanziger:2003cf}. Even more, this influence decreases when the
number of Gribov copies is large enough (cf. Tables~\ref{bc_fc},\ref{lambda_c}),
i.e. above the critical volume that we have outlined.

We wish to continue this research in a few directions: more refined 
quantities may be defined, a better accuracy is needed, larger volumes 
studied, in particular to check if the ghost propagatore dependence on the
bc/fc choice fades away in the scaling regime.  An extension 
of the study to $SU(3)$ and to unquenched gauge configurations will be
welcome.  

\paragraph{Note added in proof}
While this paper was being submitted, a preprint has 
appeared~\cite{Bogolubsky:2005wf}
which adds to the gauge group the extra
global symmetry of the quenched lattice action
with respect to the centre of the gauge group. This leads to
an extension of the notion of Gribov copies and it results
a sensitivity of the gluon propagator upon the latter.
We note that the copy dependence is mainly observed
at small lattice volumes, below the critical size found by us.

\newpage
\begin{table}[ht]
\begin{footnotesize}
\begin{center}
\begin{tabular}{r|lll|ll|ll}
\hline
$\beta$  & $\Theta_{\max_x \arrowvert \partial_\mu A^a_\mu\arrowvert}$  & $\Theta_{\text{A}_0}$ & 
$\Theta_{\delta u}$ & $P(\beta)$ & $\delta P(\beta)$ & $\langle F_\text{min}\rangle$ &  $\delta_{\text{RMS}}F_\text{min}$
\\ \hline
$2.0$ & $10^{-10}$ & $10^{-5}$ & $10^{-5}$ & $0.729620$ & $0.004738$ & $0.78416564213526$ & $0.00260692007045$
\\
$2.0$ & $10^{-14}$ & $10^{-5}$ & $10^{-9}$ & $0.729606$ & $0.004984$ & $0.78416564421249$ & $0.00260692062016$
\\
$2.0$ & $10^{-14}$ & $10^{-7}$ & $10^{-9}$ & $0.729606$ & $0.004985$ & $0.78416564421259$ & $0.00260692061867$
\\
$\rightarrow2.0$ & $10^{-18}$ & $10^{-7}$ & $10^{-9}$ & $0.729606$ & $0.004985$ & $0.78416564421258$ & $0.00260692062307$
\\
$2.0$ & $10^{-24}$ & $10^{-7}$ & $10^{-9}$ & $0.729606$ & $0.004985$ & $0.78416564421258$ & $0.00260692062150$
\\
$2.0$ & $10^{-28}$ & $10^{-7}$ & $10^{-9}$ & $0.729606$ & $0.004985$ & $0.78416564421258$ & $0.00260692062134$
\\ \hline
$2.3$ & $10^{-10}$ & $10^{-5}$ & $10^{-5}$ & $0.324636$ & $0.056660$ & $0.84466892598808$ & $0.00415444722060$
\\
$2.3$ & $10^{-14}$ & $10^{-5}$ & $10^{-9}$ & $0.324636$ & $0.056660$ & $0.84466892602737$ & $0.00415444721015$
\\
$2.3$ & $10^{-14}$ & $10^{-7}$ & $10^{-9}$ & $0.324636$ & $0.056660$ & $0.84466892602833$ & $0.00415444720619$
\\
$\rightarrow2.3$ & $10^{-18}$ & $10^{-7}$ & $10^{-9}$ & $0.324636$ & $0.056660$ & $0.84466892602843$ & $0.00415444720513$
\\
$2.3$ & $10^{-24}$ & $10^{-7}$ & $10^{-9}$ & $0.324636$ & $0.056660$ & $0.84466892602847$ & $0.00415444720356$
\\
$2.3$ & $10^{-28}$ & $10^{-7}$ & $10^{-9}$ & $0.324636$ & $0.056660$ & $0.84466892602847$ & $0.00415444720356$
\\ \hline
$2.8$ & $10^{-10}$ & $10^{-5}$ & $10^{-5}$ & $0.024946$ & $0.155566$ & $0.89393978935534$ & $0.00777035608940$
\\
$2.8$ & $10^{-14}$ & $10^{-5}$ & $10^{-9}$ & $0.024946$ & $0.155567$ & $0.89393978936422$ & $0.00777035609273$
\\
$2.8$ & $10^{-14}$ & $10^{-7}$ & $10^{-9}$ & $0.024946$ & $0.155567$ & $0.89393978936434$ & $0.00777035609111$
\\
$\rightarrow2.8$ & $10^{-18}$ & $10^{-7}$ & $10^{-9}$ & $0.024946$ & $0.155567$ & $0.89393978936448$ & $0.00777035608963$
\\
$2.8$ & $10^{-24}$ & $10^{-7}$ & $10^{-9}$ & $0.024946$ & $0.155567$ & $0.89393978936449$ & $0.00777035608962$
\\
$2.8$ & $10^{-28}$ & $10^{-7}$ & $10^{-9}$ & $0.024946$ & $0.155567$ & $0.89393978936449$ & $0.00777035608962$
\\ \hline
\end{tabular}
\end{center}
\end{footnotesize}
\caption{\footnotesize The influence of different stopping parameters (\ref{StopParameters}) on the value of
the probability $P(\beta)$. Simulations done for the  lattice size $V=8^4$, $100$ Monte-Carlo 
configurations (we uses the same set of configurations for every value of $\beta$) $\times$ $N_\text{NF}=100$ 
gauge fixings. The last two columns give the average value of $F_\text{min}$ and the standard dispersion 
of this average. The arrow indicates our choice of stopping parameters (\ref{StopParameters}).}
\label{TableStopPatameters}
\end{table}

\newpage
%
%
%
%
%
\subsection*{$SU(2)$, $V=8^4$ and $\beta=2.1$, $100$ Monte-Carlo configurations $\times$ $100$ gauge fixings}
%
\subsubsection*{Two point functions}
%
\begin{table}[!h]
\begin{footnotesize}
\centering
\begin{tabular}{c||c|c||c|c}
\hline 
$p^2$ & $F^{(2)}_{\text{fc}}(p^2)$ & $\delta F^{(2)}_{\text{fc}}(p^2)$ & $F^{(2)}_{\text{bc}}(p^2)$ & $\delta       F^{(2)}_{\text{bc}}(p^2)$ 
\\ \hline
$1$                & $4.898$ & $0.099$ & $4.687$ & $0.071$
\\ \hline
$2$                & $2.046$ & $0.039$ & $1.959$ & $0.043$
\\ \hline
$3$                &  $1.210$ & $0.021$ & $1.168$ & $0.023$
\\ \hline
$4_{~p^{[4]}=16}$  & $0.961$ & $0.023$ & $0.925$ & $0.021$
\\ \hline
$4_{~p^{[4]}=4~}$  & $0.834$ & $0.019$ & $0.801$ & $0.013$
\\ \hline
$5$                & $0.696$ & $0.007$ & $0.680$ & $0.014$
\\ \hline
\end{tabular} 
\caption{\footnotesize Ghost propagator, $V=8^4$ $\beta=2.1$}
\label{VOL_8_BETA_21_GHOST_PROPAGATOR}
\end{footnotesize}
\end{table}
%
%
\begin{table}[!h]
\begin{footnotesize}
\centering
\begin{tabular}{c||c|c||c|c}
\hline 
$p^2$ & $G^{(2)}_{\text{fc}}(p^2)$ & $\delta G^{(2)}_{\text{fc}}(p^2)$ & $G^{(2)}_{\text{bc}}(p^2)$ & $\delta       G^{(2)}_{\text{bc}}(p^2)$ 
\\ \hline
$0$                & $11.161$ & $0.438$ & $10.894$ & $0.418$
\\ \hline
$1$                & $6.225$ & $0.129$ & $6.248$ & $0.135$
\\ \hline
$2$                & $4.089$ & $0.035$ & $4.095$ & $0.043$
\\ \hline
$3$                & $2.883$ & $0.033$ & $2.868$ & $0.023$
\\ \hline
$4_{~p^{[4]}=16}$  & $2.329$ & $0.043$ & $2.305$ & $0.031$
\\ \hline
$4_{~p^{[4]}=4~}$  & $2.129$ & $0.023$ & $2.147$ & $0.015$
\\ \hline
$5$                & $1.773$ & $0.009$ & $1.785$ & $0.008$
\\ \hline
\end{tabular} 
\caption{\footnotesize Gluon propagator, $V=8^4$ $\beta=2.1$}
\label{VOL_8_BETA_21_GLUON_PROPAGATOR}
\end{footnotesize}
\end{table}
\vspace*{-1cm}
\subsubsection*{Three point functions}
%
\begin{table}[!h]
\begin{footnotesize}
\centering
\begin{tabular}{c||c|c||c|c}
\hline 
$p^2$ & $G^{(3)\text{asym}}_{\text{fc}}(p^2)$ & $\delta G^{(3)\text{asym}}_{\text{fc}}(p^2)$ & $G^{(3)\text{asym}}_{\text{bc}}(p^2)$ & $\delta       G^{(3)\text{asym}}_{\text{bc}}(p^2)$ 
\\ \hline
$1$                & $29.636$ & $2.081$ & $29.408$ & $2.721$
\\ \hline
$2$                & $19.018$ & $1.168$ & $18.387$ & $1.173$
\\ \hline
$3$                & $11.513$ & $0.971$ & $12.293$ & $1.739$
\\ \hline
$4_{~p^{[4]}=16}$  & $10.424$ & $0.857$ & $11.817$ & $1.126$
\\ \hline
$4_{~p^{[4]}=4~}$  & $6.968$ & $0.618$ & $6.547101$ & $0.471$
\\ \hline
$5$                & $5.164$ & $0.264$ & $4.697285$ & $0.293$
\\ \hline
\end{tabular} 
\caption{\footnotesize Three-gluon vertex in asymmetric kinematic configuration, $V=8^4$ $\beta=2.1$}
\label{VOL_8_BETA_21_GLUON_VERTEX_ASYM}
\end{footnotesize}
\end{table}
\begin{table}[!h]
\begin{footnotesize}
\centering
\begin{tabular}{c||c|c||c|c}
\hline 
$p^2$ & $G^{(3)\text{sym}}_{\text{fc}}(p^2)$ & $\delta G^{(3)\text{sym}}_{\text{fc}}(p^2)$ & $G^{(3)\text{sym}}_{\text{bc}}(p^2)$ & $\delta       G^{(3)\text{sym}}_{\text{bc}}(p^2)$ 
\\ \hline
$2$                & $13.613$ & $0.745$ & $14.188$ & $0.752$
\\ \hline
$4$                & $2.680$ & $0.272$ & $2.612$ & $0.272$
\\ \hline
\end{tabular} 
\caption{\footnotesize Three-gluon vertex in symmetric kinematic configuration, $V=8^4$ $\beta=2.1$}
\label{VOL_8_BETA_21_GLUON_VERTEX_SYM}
\end{footnotesize}
\end{table}
%

%
%
%
%
\newpage
%
\subsection*{$SU(2)$, $V=8^4$ and $\beta=2.2$, $100$ Monte-Carlo configurations $\times$ $100$ gauge fixings}
%
\subsubsection*{Two point functions}
%
\begin{table}[!h]\begin{footnotesize}
\centering
\begin{tabular}{c||c|c||c|c}
\hline 
$p^2$ & $F^{(2)}_{\text{fc}}(p^2)$ & $\delta F^{(2)}_{\text{fc}}(p^2)$ & $F^{(2)}_{\text{bc}}(p^2)$ & $\delta       F^{(2)}_{\text{bc}}(p^2)$ 
\\ \hline
$1$                & $4.178$ & $0.103$ & $4.100$ & $0.070$
\\ \hline
$2$                & $1.749$ & $0.045$ & $1.735$ & $0.037$
\\ \hline
$3$                & $1.042$ & $0.027$ & $1.032$ & $0.024$
\\ \hline
$4_{~p^{[4]}=16}$  & $0.835$ & $0.030$ & $0.821$ & $0.024$
\\ \hline
$4_{~p^{[4]}=4~}$  & $0.720$ & $0.016$ & $0.715$ & $0.014$
\\ \hline
$5$                & $0.614$ & $0.017$ & $0.609$ & $0.014$
\\ \hline
\end{tabular} 
\caption{\footnotesize Ghost propagator, $V=8^4$ $\beta=2.2$}
\label{VOL_8_BETA_22_GHOST_PROPAGATOR}
\end{footnotesize}\end{table}
%
%
%
\begin{table}[!h]\begin{footnotesize}
\centering
\begin{tabular}{c||c|c||c|c}
\hline 
$p^2$ & $G^{(2)}_{\text{fc}}(p^2)$ & $\delta G^{(2)}_{\text{fc}}(p^2)$ & $G^{(2)}_{\text{bc}}(p^2)$ & $\delta       G^{(2)}_{\text{bc}}(p^2)$ 
\\ \hline
$0$                & $20.170$ & $0.620$ & $20.241$ & $0.665$
\\ \hline
$1$                & $8.359$ & $0.129$ & $8.247$ & $0.122$
\\ \hline
$2$                & $4.431$ & $0.039$ & $4.433$ & $0.037$
\\ \hline
$3$                & $2.870$ & $0.028$ & $2.857$ & $0.025$
\\ \hline
$4_{~p^{[4]}=16}$  & $2.111$ & $0.038$ & $2.153$ & $0.041$
\\ \hline
$4_{~p^{[4]}=4~}$  & $2.021$ & $0.013$ & $2.038$ & $0.019$
\\ \hline
$5$                & $1.642$ & $0.014$ & $1.652$ & $0.013$
\\ 
\hline
\end{tabular} 
\caption{\footnotesize Gluon propagator, $V=8^4$ $\beta=2.2$}
\label{VOL_8_BETA_22_GLUON_PROPAGATOR}
\end{footnotesize}\end{table}
\vspace*{-1cm}
\subsubsection*{Three point functions}
%
%
\begin{table}[!h]\begin{footnotesize}
\centering
\begin{tabular}{c||c|c||c|c}
\hline 
$p^2$ & $G^{(3)\text{asym}}_{\text{fc}}(p^2)$ & $\delta G^{(3)\text{asym}}_{\text{fc}}(p^2)$ & $G^{(3)\text{asym}}_{\text{bc}}(p^2)$ & $\delta       G^{(3)\text{asym}}_{\text{bc}}(p^2)$ 
\\ \hline
$1$                & $78.913$ & $4.405$ & $81.523$ & $5.131$
\\ \hline
$2$                & $36.170$ & $1.393$ & $35.784$ & $1.809$
\\ \hline
$3$                & $17.851$ & $2.147$ & $18.475$ & $1.860$
\\ \hline
$4_{~p^{[4]}=16}$  & $20.516$ & $1.064$ & $20.621$ & $1.115$
\\ \hline
$4_{~p^{[4]}=4~}$  & $13.097$ & $0.645$ & $13.382$ & $0.595$
\\ \hline
$5$                & $8.920$ & $0.336$ & $8.985$ & $0.383$
\\ 
\hline
\end{tabular} 
\caption{\footnotesize Three-gluon vertex in asymmetric kinematic configuration, $V=8^4$ $\beta=2.2$}
\label{VOL_8_BETA_22_GLUON_VERTEX_ASYM}
\end{footnotesize}\end{table}
%
%
\begin{table}[!h]\begin{footnotesize}
\centering
\begin{tabular}{c||c|c||c|c}
\hline 
$p^2$ & $G^{(3)\text{sym}}_{\text{fc}}(p^2)$ & $\delta G^{(3)\text{sym}}_{\text{fc}}(p^2)$ & $G^{(3)\text{sym}}_{\text{bc}}(p^2)$ & $\delta       G^{(3)\text{sym}}_{\text{bc}}(p^2)$ 
\\ \hline
$2$                & $21.099$ & $1.212$ & $20.483$ & $1.148$
\\ \hline
$4$                & $2.494$ & $0.185$ & $2.512$ & $0.189$
\\ \hline
\hline
\end{tabular} 
\caption{\footnotesize Three-gluon vertex in symmetric kinematic configuration, $V=8^4$ $\beta=2.2$}
\label{VOL_8_BETA_22_GLUON_VERTEX_SYM}
\end{footnotesize}\end{table}
%
%
%
%

%
%
%
%
%
\newpage
%
\subsection*{$SU(2)$, $V=8^4$ and $\beta=2.3$, $100$ Monte-Carlo configurations $\times$ $100$ gauge fixings}
%
%
\subsubsection*{Two point functions}
%
\begin{table}[!h]\begin{footnotesize}
\centering
\begin{tabular}{c||c|c||c|c}
\hline 
$p^2$ & $F^{(2)}_{\text{fc}}(p^2)$ & $\delta F^{(2)}_{\text{fc}}(p^2)$ & $F^{(2)}_{\text{bc}}(p^2)$ & $\delta       F^{(2)}_{\text{bc}}(p^2)$ 
\\ \hline
$1$                & $3.742$ & $0.109$ & $3.656$ & $0.084$
\\ \hline
$2$                & $1.569$ & $0.043$ & $1.544$ & $0.032$
\\ \hline
$3$                & $0.945$ & $0.022$ & $0.941$ & $0.019$
\\ \hline
$4_{~p^{[4]}=16}$  & $0.755$ & $0.021$ & $0.751$ & $0.016$
\\ \hline
$4_{~p^{[4]}=4~}$  & $0.669$ & $0.015$ & $0.663$ & $0.013$
\\ \hline
$5$                & $0.562$ & $0.009$ & $0.564$ & $0.009$
\\ 
\hline
\end{tabular} 
\caption{\footnotesize Ghost propagator, $V=8^4$ $\beta=2.3$}
\label{VOL_8_BETA_23_GHOST_PROPAGATOR}
\end{footnotesize}\end{table}
%
%
%
\begin{table}[!h]\begin{footnotesize}
\centering
\begin{tabular}{c||c|c||c|c}
\hline 
$p^2$ & $G^{(2)}_{\text{fc}}(p^2)$ & $\delta G^{(2)}_{\text{fc}}(p^2)$ & $G^{(2)}_{\text{bc}}(p^2)$ & $\delta       G^{(2)}_{\text{bc}}(p^2)$ 
\\ \hline
$0$                & $35.140$ & $1.555$ & $34.659$ & $1.597$
\\ \hline
$1$                & $9.7073$ & $0.107$ & $9.839$ & $0.156$
\\ \hline
$2$                & $4.487$ & $0.054$ & $4.503$ & $0.055$
\\ \hline
$3$                & $2.668$ & $0.027$ & $2.652$ & $0.018$
\\ \hline
$4_{~p^{[4]}=16}$  & $1.934$ & $0.022$ & $1.919$ & $0.020$
\\ \hline
$4_{~p^{[4]}=4~}$  & $1.811$ & $0.016$ & $1.823$ & $0.015$
\\ \hline
$5$                & $1.414$ & $0.014$ & $1.414$ & $0.014$
\\ 
\hline
\end{tabular} 
\caption{\footnotesize Gluon propagator, $V=8^4$ $\beta=2.3$}
\label{VOL_8_BETA_23_GLUON_PROPAGATOR}
\end{footnotesize}\end{table}
\vspace*{-1cm}
\subsubsection*{Three point functions}
%
%
\begin{table}[!h]\begin{footnotesize}
\centering
\begin{tabular}{c||c|c||c|c}
\hline 
$p^2$ & $G^{(3)\text{asym}}_{\text{fc}}(p^2)$ & $\delta G^{(3)\text{asym}}_{\text{fc}}(p^2)$ & $G^{(3)\text{asym}}_{\text{bc}}(p^2)$ & $\delta       G^{(3)\text{asym}}_{\text{bc}}(p^2)$ 
\\ \hline
$1$                & $614.254$ & $42.333$ & $641.328$ & $47.008$
\\ \hline
$2$                & $165.962$ & $7.456$ & $161.354$ & $7.730$
\\ \hline
$3$                & $65.054$ & $2.786$ & $63.734$ & $2.734$
\\ \hline
$4_{~p^{[4]}=16}$  & $33.456$ & $3.787$ & $32.685$ & $3.535$
\\ \hline
$4_{~p^{[4]}=4~}$  & $33.770$ & $1.801$ & $35.883$ & $2.099$
\\ \hline
$5$                & $20.471$ & $0.9744$ & $19.994$ & $1.168$
\\ 
\hline
\end{tabular} 
\caption{\footnotesize Three-gluon vertex in asymmetric kinematic configuration, $V=8^4$ $\beta=2.3$}
\label{VOL_8_BETA_23_GLUON_VERTEX_ASYM}
\end{footnotesize}\end{table}
%
%
\begin{table}[!h]\begin{footnotesize}
\centering
\begin{tabular}{c||c|c||c|c}
\hline 
$p^2$ & $G^{(3)\text{sym}}_{\text{fc}}(p^2)$ & $\delta G^{(3)\text{sym}}_{\text{fc}}(p^2)$ & $G^{(3)\text{sym}}_{\text{bc}}(p^2)$ & $\delta       G^{(3)\text{sym}}_{\text{bc}}(p^2)$ 
\\ \hline
$2$                & $26.483$ & $1.260$ & $26.774$ & $1.280$
\\ \hline
$4$                & $2.249$ & $0.105$ & $2.424$ & $0.104$
\\ \hline
\hline
\end{tabular} 
\caption{\footnotesize Three-gluon vertex in symmetric kinematic configuration, $V=8^4$ $\beta=2.3$}
\label{VOL_8_BETA_23_GLUON_VERTEX_SYM}
\end{footnotesize}\end{table}
%
%

%
%
%
%
%
\newpage
%
\subsection*{$SU(2)$, $V=16^4$ and $\beta=2.1$, $100$ Monte-Carlo configurations $\times$ $100$ gauge fixings}
\subsubsection*{Two point functions}
%
\begin{table}[!h]\begin{footnotesize}
\centering
\begin{tabular}{c||c|c||c|c}
\hline 
$p^2$ & $F^{(2)}_{\text{fc}}(p^2)$ & $\delta F^{(2)}_{\text{fc}}(p^2)$ & $F^{(2)}_{\text{bc}}(p^2)$ & $\delta       F^{(2)}_{\text{bc}}(p^2)$ 
\\ \hline
$1$                & $23.930$ & $0.304$ & $22.615$ & $0.226$
\\ \hline
$2$                & $10.538$ & $0.200$ & $10.188$ & $0.113$
\\ \hline
$3$                & $6.372$ & $0.159$ & $6.257$ & $0.087$
\\ \hline
$4_{~p^{[4]}=16}$  & $4.551$ & $0.116$ & $4.405$ & $0.103$
\\ \hline
$4_{~p^{[4]}=4~}$  & $4.489$ & $0.119$ & $4.421$ & $0.062$
\\ \hline
$5$                & $3.437$ & $0.086$ & $3.344$ & $0.053$
\\ 
\hline
\end{tabular} 
\caption{\footnotesize Ghost propagator, $V=16^4$ $\beta=2.1$}
\label{VOL_16_BETA_21_GHOST_PROPAGATOR}
\end{footnotesize}\end{table}
%
%
\begin{table}[!h]\begin{footnotesize}
\centering
\begin{tabular}{c||c|c||c|c}
\hline 
$p^2$ & $G^{(2)}_{\text{fc}}(p^2)$ & $\delta G^{(2)}_{\text{fc}}(p^2)$ & $G^{(2)}_{\text{bc}}(p^2)$ & $\delta       G^{(2)}_{\text{bc}}(p^2)$ 
\\ \hline
$0$                & $8.485$ & $0.440$ & $8.309$ & $0.251$
\\ \hline
$1$                & $8.182$ & $0.122$ & $7.871$ & $0.102$
\\ \hline
$2$                & $7.186$ & $0.073$ & $7.022$ & $0.071$
\\ \hline
$3$                & $6.359$ & $0.037$ & $6.385$ & $0.037$
\\ \hline
$4_{~p^{[4]}=16}$  & $6.058$ & $0.091$ & $5.874$ & $0.086$
\\ \hline
$4_{~p^{[4]}=4~}$  & $5.710$ & $0.081$ & $5.669$ & $0.067$
\\ \hline
$5$                & $5.161$ & $0.029$ & $5.195$ & $0.026$
\\ 
\hline
\end{tabular} 
\caption{\footnotesize Gluon propagator, $V=16^4$ $\beta=2.1$}
\label{VOL_16_BETA_21_GLUON_PROPAGATOR}
\end{footnotesize}\end{table}
\vspace*{-1cm}
\subsubsection*{Three point functions}
%
%
\begin{table}[!h]\begin{footnotesize}
\centering
\begin{tabular}{c||c|c||c|c}
\hline 
$p^2$ & $G^{(3)\text{sym}}_{\text{fc}}(p^2)$ & $\delta G^{(3)\text{sym}}_{\text{fc}}(p^2)$ & $G^{(3)\text{sym}}_{\text{bc}}(p^2)$ & $\delta       G^{(3)\text{sym}}_{\text{bc}}(p^2)$ 
\\ \hline
$2$                & $19.861$ & $3.809$ & $27.689$ & $5.911$
\\ \hline
$4$                & $14.769$ & $2.586$ & $18.952$ & $2.937$
\\ \hline
\hline
\end{tabular} 
\caption{\footnotesize Three-gluon vertex in symmetric kinematic configuration, $V=16^4$ $\beta=2.1$}
\label{VOL_16_BETA_21_GLUON_VERTEX_SYM}
\end{footnotesize}\end{table}
%
%
%
%
%
%
\newpage
%
\subsection*{$SU(2)$, $V=16^4$ and $\beta=2.2$, $100$ Monte-Carlo configurations $\times$ $100$ gauge fixings}
%
%
\subsubsection*{Two point functions}
%
\begin{table}[!h]\begin{footnotesize}
\centering
\begin{tabular}{c||c|c||c|c}
\hline 
$p^2$ & $F^{(2)}_{\text{fc}}(p^2)$ & $\delta F^{(2)}_{\text{fc}}(p^2)$ & $F^{(2)}_{\text{bc}}(p^2)$ & $\delta       F^{(2)}_{\text{bc}}(p^2)$ 
\\ \hline
$1$                &  $20.867$ & $0.376$ & $20.406$ & $0.392$
\\ \hline
$2$                & $9.151$ & $0.145$ & $9.070$ & $0.192$
\\ \hline
$3$                & $5.499$ & $0.083$ & $5.460$ & $0.085$
\\ \hline
$4_{~p^{[4]}=16}$  & $3.861$ & $0.095$ & $3.838$ & $0.082$
\\ \hline
$4_{~p^{[4]}=4~}$  & $3.803$ & $0.067$ & $3.819$ & $0.068$
\\ \hline
$5$                & $2.929$ & $0.041$ & $2.979$ & $0.067$
\\ 
\hline
\end{tabular} 
\caption{\footnotesize Ghost propagator, $V=16^4$ $\beta=2.2$}
\label{VOL_16_BETA_22_GHOST_PROPAGATOR}
\end{footnotesize}\end{table}
%
%
\begin{table}[!h]\begin{footnotesize}
\centering
\begin{tabular}{c||c|c||c|c}
\hline 
$p^2$ & $G^{(2)}_{\text{fc}}(p^2)$ & $\delta G^{(2)}_{\text{fc}}(p^2)$ & $G^{(2)}_{\text{bc}}(p^2)$ & $\delta       G^{(2)}_{\text{bc}}(p^2)$ 
\\ \hline
$0$                & $14.473$ & $0.676$ & $15.380$ & $0.635$
\\ \hline
$1$                & $12.614$ & $0.255$ & $12.330$ & $0.184$
\\ \hline
$2$                & $10.564$ & $0.066$ & $10.531$ & $0.097$
\\ \hline
$3$                & $8.813$ & $0.081$ & $8.769$ & $0.061$
\\ \hline
$4_{~p^{[4]}=16}$  & $7.760$ & $0.162$ & $7.577$ & $0.089$
\\ \hline
$4_{~p^{[4]}=4~}$  & $7.447$ & $0.052$ & $7.429$ & $0.073$
\\ \hline
$5$                & $6.393$ & $0.052$ & $6.395$ & $0.044$
\\ 
\hline
\end{tabular} 
\caption{\footnotesize Gluon propagator, $V=16^4$ $\beta=2.2$}
\label{VOL_16_BETA_22_GLUON_PROPAGATOR}
\end{footnotesize}\end{table}
\vspace*{-1cm}
\subsubsection*{Three point functions}
%
%
\begin{table}[!h]\begin{footnotesize}
\centering
\begin{tabular}{c||c|c||c|c}
\hline 
$p^2$ & $G^{(3)\text{asym}}_{\text{fc}}(p^2)$ & $\delta G^{(3)\text{asym}}_{\text{fc}}(p^2)$ & $G^{(3)\text{asym}}_{\text{bc}}(p^2)$ & $\delta       G^{(3)\text{asym}}_{\text{bc}}(p^2)$ 
\\ \hline
$1$                & $136.266$ & $91.211$ & $115.376$ & $53.064$
\\ \hline
$2$                & $93.912$ & $39.689$ & $90.432$ & $26.263$
\\ \hline
$3$                & $84.119$ & $17.125$ & $81.880$ & $17.916$
\\ \hline
$4_{~p^{[4]}=16}$  & $101.745$ & $17.120$ & $99.140$ & $18.988$
\\ \hline
$4_{~p^{[4]}=4~}$  & $78.176$ & $15.640$ & $63.382$ & $14.634$
\\ \hline
$5$                & $51.954$ & $7.666$ & $56.088$ & $8.017$
\\ 
\hline
\end{tabular} 
\caption{\footnotesize Three-gluon vertex in asymmetric kinematic configuration, $V=16^4$ $\beta=2.2$}
\label{VOL_16_BETA_22_GLUON_VERTEX_ASYM}
\end{footnotesize}\end{table}
%
%
\begin{table}[!h]\begin{footnotesize}
\centering
\begin{tabular}{c||c|c||c|c}
\hline 
$p^2$ & $G^{(3)\text{sym}}_{\text{fc}}(p^2)$ & $\delta G^{(3)\text{sym}}_{\text{fc}}(p^2)$ & $G^{(3)\text{sym}}_{\text{bc}}(p^2)$ & $\delta       G^{(3)\text{sym}}_{\text{bc}}(p^2)$ 
\\ \hline
$2$                & $93.161$ & $11.297$ & $87.022$ & $10.680$
\\ \hline
$4$                & $34.756$ & $7.434$ & $37.655$ & $5.797$
\\ \hline
\hline
\end{tabular} 
\caption{\footnotesize Three-gluon vertex in symmetric kinematic configuration, $V=16^4$ $\beta=2.2$}
\label{VOL_16_BETA_22_GLUON_VERTEX_SYM}
\end{footnotesize}\end{table}
%
%
%
%
%
%
%
\newpage
%
\subsection*{$SU(2)$, $V=16^4$ and $\beta=2.3$, $100$ Monte-Carlo configurations $\times$ $100$ gauge fixings}
%
%
\subsubsection*{Two point functions}
%
\begin{table}[!h]\begin{footnotesize}
\centering
\begin{tabular}{c||c|c||c|c}
\hline 
$p^2$ & $F^{(2)}_{\text{fc}}(p^2)$ & $\delta F^{(2)}_{\text{fc}}(p^2)$ & $F^{(2)}_{\text{bc}}(p^2)$ & $\delta       F^{(2)}_{\text{bc}}(p^2)$ 
\\ \hline
$1$                & $18.326$ & $0.297$ & $17.157$ & $0.172$
\\ \hline
$2$                & $7.993$ & $0.150$ & $7.645$ & $0.089$
\\ \hline
$3$                & $4.824$ & $0.086$ & $4.684$ & $0.060$
\\ \hline
$4_{~p^{[4]}=16}$  & $3.431$ & $0.053$ & $3.317$ & $0.052$
\\ \hline
$4_{~p^{[4]}=4~}$  & $3.364$ & $0.054$ & $3.338$ & $0.048$
\\ \hline
$5$                & $2.581$ & $0.025$ & $2.533$ & $0.036$
\\ 
\hline
\end{tabular} 
\caption{\footnotesize Ghost propagator, $V=16^4$ $\beta=2.3$}
\label{VOL_16_BETA_23_GHOST_PROPAGATOR}
\end{footnotesize}\end{table}
%
%
\begin{table}[!h]\begin{footnotesize}
\centering
\begin{tabular}{c||c|c||c|c}
\hline 
$p^2$ & $G^{(2)}_{\text{fc}}(p^2)$ & $\delta G^{(2)}_{\text{fc}}(p^2)$ & $G^{(2)}_{\text{bc}}(p^2)$ & $\delta       G^{(2)}_{\text{bc}}(p^2)$ 
\\ \hline
$0$                & $32.705$ & $1.063$ & $31.557$ & $0.721$
\\ \hline
$1$                & $21.940$ & $0.444$ & $21.616$ & $0.385$
\\ \hline
$2$                & $15.007$ & $0.122$ & $15.206$ & $0.166$
\\ \hline
$3$                & $11.400$ & $0.107$ & $11.235$ & $0.116$
\\ \hline
$4_{~p^{[4]}=16}$  & $8.848$ & $0.190$ & $8.882$ & $0.108$
\\ \hline
$4_{~p^{[4]}=4~}$  & $8.904$ & $0.118$ & $8.782$ & $0.099$
\\ \hline
$5$                & $7.057$ & $0.062$ & $6.978$ & $0.045$
\\ 
\hline
\end{tabular} 
\caption{\footnotesize Gluon propagator, $V=16^4$ $\beta=2.3$}
\label{VOL_16_BETA_23_GLUON_PROPAGATOR}
\end{footnotesize}\end{table}
\vspace*{-1cm}
\subsubsection*{Three point functions}
%
%
\begin{table}[!h]\begin{footnotesize}
\centering
\begin{tabular}{c||c|c||c|c}
\hline 
$p^2$ & $G^{(3)\text{asym}}_{\text{fc}}(p^2)$ & $\delta G^{(3)\text{asym}}_{\text{fc}}(p^2)$ & $G^{(3)\text{asym}}_{\text{bc}}(p^2)$ & $\delta       G^{(3)\text{asym}}_{\text{bc}}(p^2)$ 
\\ \hline
$1$                & $1192.828$ & $250.675$ & $779.501$ & $204.768$
\\ \hline
$2$                & $566.632$ & $39.161$ & $505.435$ & $46.025$
\\ \hline
$3$                & $402.454$ & $32.415$ & $388.260$ & $26.754$
\\ \hline
$4_{~p^{[4]}=16}$  & $272.205$ & $35.491$ & $242.150$ & $30.928$
\\ \hline
$4_{~p^{[4]}=4~}$  & $239.204$ & $35.152$ & $241.295$ & $25.467$
\\ \hline
$5$                & $174.210$ & $14.126$ & $169.739$ & $10.825$
\\ \hline
$6$                & $130.209$ & $7.523$ & $125.733$ & $7.563$
\\ 
\hline
\end{tabular} 
\caption{\footnotesize Three-gluon vertex in asymmetric kinematic configuration, $V=16^4$ $\beta=2.3$}
\label{VOL_16_BETA_23_GLUON_VERTEX_ASYM}
\end{footnotesize}\end{table}
%
%
\begin{table}[!h]\begin{footnotesize}
\centering
\begin{tabular}{c||c|c||c|c}
\hline 
$p^2$ & $G^{(3)\text{sym}}_{\text{fc}}(p^2)$ & $\delta G^{(3)\text{sym}}_{\text{fc}}(p^2)$ & $G^{(3)\text{sym}}_{\text{bc}}(p^2)$ & $\delta       G^{(3)\text{sym}}_{\text{bc}}(p^2)$ 
\\ \hline
$2$                & $283.901$ & $15.417$ & $313.824$ & $15.595$
\\ \hline
$4$                & $81.695$ & $5.785$ & $72.573$ & $5.787$
\\ \hline
\hline
\end{tabular} 
\caption{\footnotesize Three-gluon vertex in symmetric kinematic configuration, $V=16^4$ $\beta=2.3$}
\label{VOL_16_BETA_23_GLUON_VERTEX_SYM}
\end{footnotesize}\end{table}
%
%
\newpage
\bibliographystyle{hunsrt.bst}
\bibliography{references}
\end{document}